\documentclass[preprint,showpacs,showkeys,preprintnumbers,amsmath,amssymb]{revtex4}

\usepackage{dcolumn}
\usepackage{bm}

\begin{document}

\title{Exact and quasiclassical Green's functions of two-dimensional electron
gas with Rashba-Dresselhaus spin-orbit interaction in parallel
magnetic field}


\author{I. V. Kozlov}
\email{kozlov@ilt.kharkov.ua}
\author{Yu. A. Kolesnichenko}
\email{kolesnichenko@ilt.kharkov.ua}
\affiliation{B. Verkin
Institute for Low Temperature Physics and Engineering of the
National Academy of Sciences of Ukraine, 47 Nauky Ave., Kharkiv
61103, Ukraine}

\begin{abstract}
New exact and asymptotical results for the one particle Green's
function of 2D electrons with combined Rashba - Dresselhaus spin -
orbit interaction in the presence of in-plane uniform magnetic
field are presented. A special case that allows an exact
analytical solution is also highlighted. To demonstrate the
advantages of our approach we apply the obtained Green's function
to calculation of electron density and magnetization.
\end{abstract}



\maketitle

\section{\label{sec:level1} Introduction}

One particle Green's functions (GF) is widely used in quantum
mechanics and solid state physics \cite{bib1,bib2}. Although the
simple analytical formulas for GFs of the free electron gas can be
found in some textbooks (see, for example, \cite{bib2}), in the
presence of interactions and in external fields in most cases only
complicated integral representations of GF are available. Normaly
using the GF in the form of a multiple integral is extremely
uncomfortable both for analytical analysis and for numerical
computations. Therefore the search of exact, simplified and
asymptotic results for GFs attracts the constant attention of
theorists and mathematicians (see, for example,
\cite{bib3,bib4,bib5}).

In last decade the spintronics development has triggered off
investigations of two-dimensional (2D) electron systems with
spin-orbit interaction (SOI) (for review see \cite{bib6,bib7}).
Particularly the 2D systems with combined Rashba \cite{bib8} and
Dresselhaus \cite{bib9} SOI possesses new perspective properties
(for review see \cite{bib10}). The GF for this case without
magnetic field and its asymptotic behavior have been discussed in
the papers \cite{bib11,bib12}. Explicit GF for each Rashba and
Dresselhaus spin-orbit Hamiltonians with uniform perpendicular
magnetic field have been derived in Ref. \cite{bib3}.

In this paper we obtain some exact and asymptotic expressions at
zero temperature for the time independent GF of 2D electron gas
with combined Rashba - Dresselhaus SOI for arbitrary values of SOI
constants and arbitrary uniform magnetic field strength and
direction parallel to the plane of the conductor.  The structure
of this paper is as it follows. In Sec. 2 we discuss the
Hamiltonian of the system, its eigenvalues and eigenfunctions. The
GF in coordinate space is presented as the sum of two parts
describing separated contributions of two spin - orbit splitted
branches of electron energy spectrum.  In Sec. 3 we reduce the
general expression for the GF to single integral of well-known
special functions. The obtained  formula is valid at arbitrary
values of parameters. In Sec. 4 we derive asymptotic formulas for
GF for large value of coordinate variable. In Sec. 5 we find the
exact GF for the special case of equal SOI constants and certain
direction of magnetic field. In Sec.6 as a demonstration we apply
our results for calculation of electron density of states and
density of magnetization. We conclude the paper with the final
remarks and summing up main results in Sec.7.

\section{Basic formulas}

Let us consider the Hamiltonian of a two-dimensional (2D)
noninteracting electron gas with Rashba and Dresselhaus SOI in the
magnetic field $ B=\left( {{B}_{x}},{{B}_{y}},0 \right) $.   Using
the Coulomb gauge $ \mathbf{A}=\left( 0,0,{{B}_{x}}y-{{B}_{y}}x
\right) $ , $ \nabla \cdot \mathbf{A}=0 $, we write 2D Hamiltonian
of the system as the sum of four terms
\begin{equation}
\hat{H}={{\hat{H}}_{0}}+{{\hat{H}}_{R}}+{{\hat{H}}_{D}}+{{\hat{H}}_{B}}.
\label{eq1}
\end{equation}
Here $ {{\hat{H}}_{0}}=\frac{{{\hbar }^{2}}\left(
\hat{K}_{x}^{2}+\hat{K}_{y}^{2} \right)}{2m}{{\sigma }_{0}} $ is
the Hamiltonian of 2D free electron gas, $ {{\hat{H}}_{R}}=\alpha
({{\sigma }_{x}}{{\hat{K}}_{y}}-{{\sigma }_{y}}{{\hat{K}}_{x}}) $
and ${{\hat{H}}_{D}}=\beta ({{\sigma
}_{x}}{{\hat{K}}_{x}}-{{\sigma }_{y}}{{\hat{K}}_{y}})$ are
Hamiltonians of Rashba and Dresselhaus SOI, respectively,
${{\hat{H}}_{B}}=\frac{{{g}^{*}}}{2}{{\mu }_{B}}\left(
{{B}_{x}}{{\sigma }_{x}}+{{B}_{y}}{{\sigma }_{y}} \right)$ is the
Hamiltonian of interaction between electron spin and magnetic
field, $ {{\hat{K}}_{x,y}}=-i{{\nabla }_{x,y}} $ is the wave
vector operator, $m$ is effective electron mass, ${{\sigma
}_{x,y,z}}$ are Pauli matrices, ${{\hat{\sigma }}_{0}}$ is unit
matrix $2\times 2$, $\alpha $ and $\beta $ are Rashba ($\alpha $)
and Dresselhaus$ \left( \beta  \right) $constants of SOI, $ {{\mu
}_{B}} $ is the Bohr magneton, ${{g}^{*}}$ is an effective
g-factor of the  2D system. We rewrite the total Hamiltonian
(\ref{eq1}) in the following form
\begin{equation}
\hat{H}={{\hat{H}}_{0}}+\mathbf{\hat{R}}  \mathbf{\sigma },
\label{eq2}
\end{equation}
where $\mathbf{\sigma }=\left( {{\sigma }_{x}},{{\sigma
}_{y}},{{\sigma }_{z}} \right)$ is Pauli vector,
\begin{equation}
{{\hat{R}}_{x}}={{h}_{x}}+\alpha {{\hat{K}}_{y}}+\beta
{{\hat{K}}_{x}};\quad {{\hat{R}}_{y}}={{h}_{y}}-\alpha
{{\hat{K}}_{x}}-\beta {{\hat{K}}_{y}},\quad
{{\hat{R}}_{z}}=0;\quad {{h}_{x,y}}=\frac{{{g}^{*}}}{2}{{\mu
}_{B}}{{B}_{x,y}}. \label{eq3}
\end{equation}

The eigenvalues and the eigenfunctions of the Hamiltonian
(\ref{eq2}) are (see, for example,\cite{bib13})
\begin{equation}
{{\epsilon }_{1,2}}\left( \mathbf{k} \right)={{\epsilon }_{0}}\pm
R\left( {{k}_{x}},{{k}_{y}} \right);
\label{eq4}
\end{equation}
\begin{equation}
\psi _{1,2}^{{}}\left( \mathbf{r}
\right)=\frac{1}{2\pi \sqrt{2}}{{e}^{i\mathbf{kr}}}\left(
\begin{aligned}
  & 1 \\
 & {{e}^{i{{\theta }_{1,2}}}} \\
\end{aligned} \right)\equiv \frac{1}{2\pi }{{e}^{i\mathbf{kr}}}\phi \left( {{\theta }_{1,2}} \right).
\label{eq5}
\end{equation}
We introduce notations:
\begin{equation}
sin{{\theta }_{1}}=\frac{{{R}_{y}}}{R};\quad cos{{\theta
}_{1}}=\frac{{{R}_{x}}}{R},\quad {{\theta }_{2}}={{\theta
}_{1}}+\pi ,
\label{eq6}
\end{equation}
\begin{equation}
\mathbf{k}=\left( {{k}_{x}},{{k}_{y}},0 \right),\quad {{\epsilon
}_{0}}=\frac{{{\hbar }^{2}}{{k}^{2}}}{2m},\quad
k=\sqrt{k_{x}^{2}+k_{y}^{2}},
\label{eq7}
\end{equation}
\begin{equation}
R=\sqrt{R_{x}^{2}+R_{y}^{2}}=\sqrt{{{\left( {{h}_{x}}+\alpha
{{k}_{y}}+\beta {{k}_{x}} \right)}^{2}}+{{\left( {{h}_{y}}-\alpha
{{k}_{x}}-\beta {{k}_{y}} \right)}^{2}}}.
\label{eq8}
\end{equation}
The angles $ {{\theta }_{1,2}} $ define the average spin direction
for two branches of energy spectrum  (\ref{eq4})
\begin{equation}
{{\mathbf{s}}_{1,2}}\left( \theta  \right)={{\phi }^{\dagger
}}\left( {{\theta }_{1,2}} \right)\mathbf{\sigma }\phi \left(
{{\theta }_{1,2}} \right) = \left( \cos {{\theta
}_{1,2}},sin{{\theta }_{1,2}},0 \right). \label{eq9}
\end{equation}
The electron GF corresponding Hamiltonian (\ref{eq2}) for complex
$\epsilon $ in coordinate representation can be written as
\begin{equation}
\begin{aligned}
  & \hat{G}\left( \epsilon ,\mathbf{r} \right)=\frac{1}{{{(2\pi )}^{2}}}{{\int\limits_{-\infty }^{\infty }
  \frac{d\mathbf{k} {e^{i\mathbf{kr}}}}{ \left( \epsilon -{{\epsilon }_{0}} \right){{\sigma }_{0}}-\mathbf{R\sigma }}
  }}
= \\
 & \frac{1}{2{{(2\pi )}^{2}}}\sum\limits_{j=1,2}{\int\limits_{-\infty }^{\infty }{d\mathbf{k}}}\frac{{{e}^{i\mathbf{kr}}}}{\epsilon -{{\epsilon }_{j}}}\left( {{\sigma }_{0}}+{{\sigma }_{x}}\cos {{\theta }_{j}}+{{\sigma }_{y}}\sin {{\theta }_{j}} \right),\quad \epsilon \in \mathbb{C}, \\
\end{aligned}
\label{eq10}
\end{equation}
We used Eqs. (\ref{eq4}), (\ref{eq6}) and the identities $
{{\epsilon }_{1,2}}-{{\epsilon }_{0}}=\pm R $,
\begin{equation}
\begin{aligned}
  & \left( \left( \epsilon -{{\epsilon }_{0}} \right){{\sigma }_{0}}-\mathbf{R\sigma } \right)\left( \left( \epsilon -{{\epsilon }_{0}} \right){{\sigma }_{0}}+\mathbf{R\sigma } \right)= \\
 & \left[ {{\left( \epsilon -{{\epsilon }_{0}} \right)}^{2}}-{{R}^{2}} \right]{{\sigma }_{0}}=\left( \epsilon -{{\epsilon }_{1}} \right)\left( \epsilon -{{\epsilon }_{2}} \right){{\sigma }_{0}}; \\
\end{aligned}
\label{eq11}
\end{equation}
\begin{equation}
\begin{aligned}
  & \frac{\left( \left( \epsilon -{{\epsilon }_{0}} \right){{\sigma }_{0}}+\mathbf{R\sigma } \right)}{\left( \left( \epsilon -{{\epsilon }_{0}} \right){{\sigma }_{0}}-\mathbf{R\sigma } \right)\left( \left( \epsilon -{{\epsilon }_{0}} \right){{\sigma }_{0}}+\mathbf{R\sigma } \right)}= \\
 & \frac{1}{2R}\left\{ \frac{\left( {{\epsilon }_{1}}-{{\epsilon }_{0}} \right){{\sigma }_{0}}+\mathbf{R\sigma }}{\epsilon -{{\epsilon }_{1}}}-\frac{\left( {{\epsilon }_{2}}-{{\epsilon }_{0}} \right){{\sigma }_{0}}+\mathbf{R\sigma }}{\epsilon -{{\epsilon }_{2}}} \right\}. \\
\end{aligned}
\label{eq12}
\end{equation}
In the Eq. (\ref{eq10}) GF splits up into two independent parts
describing separate contributions of every branch of energy
spectrum (\ref{eq4}).

\section{Exact results for Green's function}

For $ \alpha \ne \beta $ we introduce new variables of integration
$ \tilde{k}, f $ as follows
\begin{equation}
{{k}_{x}}=k_{x0}^{{}}+\tilde{k}\cos f,\quad
{{k}_{y}}=k_{y0}^{{}}+\tilde{k}\sin f,
\label{eq13}
\end{equation}
\begin{equation}
{{k}_{x0}}=\frac{\alpha {{h}_{y}}+\beta {{h}_{x}}}{{{\alpha
}^{2}}-{{\beta }^{2}}};\quad {{k}_{y0}}=-\frac{\alpha
{{h}_{x}}+\beta {{h}_{y}}}{{{\alpha }^{2}}-{{\beta }^{2}}},
\label{eq14}
\end{equation}
where $ {\bf k_0} = ( k_{x0} , k_{y0} ) $ is  the point of branch
touch (see, for example Ref. \cite{bib14}). In coordinates
(\ref{eq13}) spin angles (\ref{eq6}) depend only on the wave
vector direction, the angle f, and SOI constants
\begin{equation}
\sin {{\theta }_{1,2}}\left( f \right)=\mp \frac{\alpha \cos
f+\beta \sin f}{\sqrt{{{\alpha }^{2}}+{{\beta }^{2}}+2\alpha \beta
\sin 2f}};\quad \cos {{\theta }_{1,2}}\left( f \right)=\pm
\frac{\alpha \sin f+\beta \cos f}{\sqrt{{{\alpha }^{2}}+{{\beta
}^{2}}+2\alpha \beta \sin 2f}}.
\label{eq15}
\end{equation}
It can be shown that spin direction is symmetric with respect to
the center  $ \theta_{1,2} ( f + \pi) = \theta_{1,2} ( f ) + \pi
$, indicating a convenience of  Eq. (\ref{eq13}).   The values
${{k}_{x0}},{{k}_{y0}}$ have been found out of the system of
equations $ {{R}_{x}}\left( {{k}_{x0}},{{k}_{y0}} \right)=0,
\qquad {{R}_{y}}\left( {{k}_{x0}},{{k}_{y0}} \right)=0 $. In a
shifted polar coordinates $\tilde{k},f$ (\ref{eq13})  the energies
$ {{\epsilon }_{1,2}} $ take the form
\begin{equation}
{{\epsilon }_{1,2}}\left( \tilde{k},\tilde{f}
\right)=\frac{{{\hbar }^{2}}{{{\tilde{k}}}^{2}}}{2m}-\frac{{{\hbar
}^{2}}\tilde{k}}{m}{{\lambda }_{1,2}}\left( f \right)+{{E}_{0}}.
\label{eq16}
\end{equation}
Here
\begin{equation}
{{\lambda }^{\left( 1,2 \right)}}\left( f \right)=-k_{x0}^{{}}\cos
f-k_{y0}^{{}}\sin f\mp \frac{m}{{{\hbar }^{2}}}\sqrt{{{\alpha
}^{2}}+{{\beta }^{2}}+2\alpha \beta \sin (2f)};
\label{eq17}
\end{equation}
\begin{equation}
{{E}_{0}}=\frac{{{h}^{2}}({{\alpha }^{2}}+{{\beta }^{2}})+4\alpha
\beta {{h}_{x}}{{h}_{y}}}{2m{{\hbar }^{2}}{{({{\alpha
}^{2}}-{{\beta }^{2}})}^{2}}}.
\label{eq18}
\end{equation}
The Eq.(\ref{eq16}) make it possible to write the poles $
{{\epsilon }_{1,2}}\left( k_{\pm }^{\left( 1,2 \right)},f
\right)=\epsilon $ of the Green function (\ref{eq10}) in a simple
form
\begin{equation}
k_{\pm }^{\left( 1,2 \right)}={{\lambda }^{\left( 1,2 \right)}}\pm
\sqrt{{{\xi }^{\left( 1,2 \right)}}}.
\label{eq19}
\end{equation}
\begin{equation}
{{\xi }^{\left( 1,2 \right)}}={{\left( {{\lambda }^{\left( 1,2
\right)}} \right)}^{2}}+\frac{2m(\epsilon -{{E}_{0}})}{{{\hbar
}^{2}}}.
\label{eq20}
\end{equation}
By using the roots (\ref{eq19}) in coordinates  (\ref{eq13}) one
can write a simple expansion of the  functions $ {{ \left(
\epsilon -{{\epsilon }_{1,2}} \right) }^{-1}} $
\begin{equation}
\frac{1}{\epsilon -{{\epsilon }_{1,2}}}=-\frac{2m}{{{\hbar
}^{2}}\tilde{k}}\sum\limits_{\pm }{\frac{k_{\pm }^{\left( 1,2
\right)}}{k_{\pm }^{\left( 1,2 \right)}-k_{\mp }^{\left( 1,2
\right)}}\frac{1}{\tilde{k}-k_{\pm }^{\left( 1,2 \right)}}},
\label{eq21}
\end{equation}
By means of identity (\ref{eq21}21)  the GF (\ref{eq10})  is
written as the double sum
\begin{equation}
\begin{aligned}
  & G(\epsilon ,\mathbf{r})=-\frac{m}{{{\left( 2\pi \hbar  \right)}^{2}}}\exp \left[ i({{k}_{x0}}\cos {{\varphi }_{r}}+{{k}_{y0}}sin{{\varphi }_{r}})r \right]\times  \\
 & \sum\limits_{j=1,2}{\oint{d}\tilde{f}\left( 1+{{\sigma }_{x}}\cos {{\theta }_{j}}+{{\sigma }_{y}}\sin {{\theta }_{j}} \right)}\sum\limits_{\pm }{\frac{\tilde{k}_{\pm }^{\left( j \right)}}{\tilde{k}_{\pm }^{\left( j \right)}-\tilde{k}_{\mp }^{\left( j \right)}}}\int\limits_{0}^{\infty }{\frac{d\tilde{k}}{\tilde{k}-\tilde{k}_{\pm }^{\left( j \right)}}}{{e}^{i\tilde{k}r\cos (\tilde{f}-\varphi )}}\quad \epsilon \in \mathbb{C}. \\
\end{aligned}
\label{eq22}
\end{equation}
At $ r \neq 0 $ the integral  over $ \tilde{k} $ in Eq.
(\ref{eq22}) can be expressed by means of trigonometric integral
functions (see, for example \cite{bib15}),
\begin{equation}
F\left( {{k}_{0}},r \right)=\int\limits_{0}^{\infty
}{\frac{d\tilde{k}}{\tilde{k}-{{k}_{0}}}}{{e}^{i\tilde{k}r}}={{e}^{i{{k}_{0}}r}}\left[
-Ci\left( -{{k}_{0}}\left| r \right| \right)+iSi\left( {{k}_{0}}r
\right)+\frac{i\pi }{2} sign \, r \, \right],\quad r\in
\mathbb{R}, \label{eq23}
\end{equation}
where $ Si\left( z \right) $ and $ Ci\left( z \right) $ are sine
and cosine integrals,
\begin{equation}
\begin{aligned}
  & Ci\left( z \right)=-\int\limits_{z}^{\infty }{\frac{\cos z}{z}};\quad Si\left( z \right)=-\int\limits_{z}^{\infty }{\frac{\sin z}{z}}+\frac{\pi }{2};\quad z\in \mathbb{C};\quad \left| \arg z \right|<\pi ; \\
 & \underset{\delta \to +0}{\mathop{\lim }}\,Ci\left( x\pm i\delta  \right)=Ci\left( \left| x \right| \right)\pm \pi \Theta \left( -x \right);\quad x\in \mathbb{R}. \\
\end{aligned}
\label{eq24}
\end{equation}
The Eq.(\ref{eq23}) is valid for any $ {{k}_{0}}\in \mathbb{C} $
excepting semiaxis $ \operatorname{Im}{{k}_{0}}=0,\quad
0<\operatorname{Re}{{k}_{0}}<\infty $, for which usually one
introduces the retarded and advanced Green functions $
{{G}^{R\left( A \right)}}(E,\mathbf{r}) $ as the limit
\begin{equation}
{{G}^{R\left( A \right)}}(E,\mathbf{r})=\underset{\delta \to
+0}{\mathop{\lim }}\,G\left( E\pm i\delta ,\mathbf{r}
\right),\quad E\in \mathbb{R}.
\label{eq25}
\end{equation}
Using Eqs. (\ref{eq21}), (\ref{eq23}) one can rewrite the Green
function (\ref{eq10}) as
\begin{equation}
\begin{aligned}
  & G(\epsilon ,\mathbf{r})=-\frac{m}{{{\left( 2\pi \hbar  \right)}^{2}}}\exp \left[ i({{k}_{x0}}\cos {{\varphi }_{r}}+{{k}_{y0}}sin{{\varphi }_{r}})r \right]\times  \\
 & \sum\limits_{j=1,2}{\oint{d}f}\left( 1+{{\sigma }_{x}}\cos {{\theta }_{j}}+{{\sigma }_{y}}\sin {{\theta }_{j}} \right)\sum\limits_{\pm }{\frac{k_{\pm }^{\left( j \right)}}{k_{\pm }^{\left( j \right)}-k_{\mp }^{\left( j \right)}}}F\left( k_{\pm }^{\left( j \right)},r\cos (f-{{\varphi }_{r}}) \right),\quad \epsilon \in \mathbb{C}, \\
\end{aligned}
\label{eq26}
\end{equation}
where angle  $ {{\varphi }_{r}} $ defines a direction of
coordinate $ \mathbf{r}=r\left( cos{{\varphi }_{r}}, \;
sin{{\varphi }_{r}}, \; 0 \right) $. The formula (\ref{eq26})
turns to the result of the Ref. \cite{bib11} for the zero magnetic
field. The obtained GF (\ref{eq26}) is suitable for numerical
calculation under arbitrary values of all parameters. It also
gives analytical formulas in the quasiclassical case $ r \to
\infty $.

For equal SOI constants, $ \alpha =\beta $, one cannot use
coordinates (\ref{eq13}) and we introduce usual polar coordinates
\begin{equation}
{{k}_{x}}=k\cos {{\varphi }_{k}};\quad {{k}_{y}}=k\sin {{\varphi
}_{k}}.
\label{eq27}
\end{equation}
After transformations similar to those performed above we find
\begin{equation}
\hat{G}\left( \epsilon ,\mathbf{r} \right)=\frac{{{m}^{2}}}{{{\pi
}^{2}}{{\hbar }^{4}}}\sum\limits_{n=1}^{4}{\oint{d{{\varphi
}_{k}}}}\left[ \left( \epsilon -\frac{{{\hbar }^{2}}k_{n}^{2}}{2m}
\right){{\sigma }_{0}}+{{L}_{x}}\left( {{k}_{n}} \right){{\sigma
}_{x}}+{{L}_{y}}\left( {{k}_{n}} \right){{\sigma }_{y}}
\right]\frac{{{k}_{n}}F\left( {{k}_{n}},r\cos ({{\varphi
}_{k}}-{{\varphi }_{r}}) \right)}{{{\left. \partial Q/\partial k
\right|}_{k={{k}_{n}}}}},
\label{eq28}
\end{equation}
where $ {{k}_{n}} $ are the roots of quartic polynomial equation
without cubic term
\begin{equation}
Q\left( {{k}_{n}},{{\varphi }_{k}} \right)=0,
\label{eq29}
\end{equation}
\begin{equation}
\begin{aligned}
  & Q\left( k,{{\varphi }_{k}} \right)={{k}^{4}}-{{k}^{2}}\frac{4m}{{{\hbar }^{2}}}\left[ \epsilon +\frac{2m{{\alpha }^{2}}}{{{\hbar }^{2}}}\left( 1+\sin 2{{\varphi }_{k}} \right) \right]- \\
 & k\frac{8{{m}^{2}}\alpha h}{{{\hbar }^{4}}}\left[ \sin \left( {{\varphi }_{k}}-{{\varphi }_{h}} \right)+\cos \left( {{\varphi }_{k}}+{{\varphi }_{h}} \right) \right]+\frac{4{{m}^{2}}}{{{\hbar }^{4}}}\left( {{\epsilon }^{2}}-{{h}^{2}} \right), \\
\end{aligned}
\label{eq30}
\end{equation}
\begin{equation}
{{L}_{x}}=\left( h\cos {{\varphi }_{h}}+\sqrt{2}\alpha k\sin
\left( {{\varphi }_{k}}+\frac{\pi }{4} \right) \right),\quad
{{L}_{y}}=\left( h\sin {{\varphi }_{h}}-\sqrt{2}\alpha k\sin
\left( {{\varphi }_{k}}+\frac{\pi }{4} \right) \right),
\label{eq31}
\end{equation}
the function $ F\left( k,r \right) $ is given by Eq. (\ref{eq23}),
$ {{\varphi }_{h}} $ defines the magnetic field direction, $
\mathbf{h}=h\left( \cos {{\varphi }_{h}},\sin {{\varphi }_{h}},0
\right) $. Though the Eq. (\ref{eq29}) has exact analytical
solutions (see, for example,\cite{bib16}) they are very
complicated and not suitable for analytical calculation.
Nevertheless the Eq. (\ref{eq28}) may be pracically convinient in
numerical analysis. As for the particular case of magnetic field
direction along the symmetry axis $ {{k}_{x}}=-{{k}_{y}} $ we
consider it in Sec.5. and the GF has been expressed by means of
Bessel functions.

\section{Quasiclassical Green's function}

Quasiclassical  approximation can be applied in physical
investigations, if characteristic length scales of the problem are
much larger than Fermi wavelength $ {{\lambda }_{F}} $ which has
of the order of inverse wave vector $ {{k}^{-1}} $ at Fermi level.
Since the GF  oscillates as a function of the coordinate $r$ on a
scale $ r\sim {{k}^{-1}} $ in framework of quasiclassical
approximation in most cases the asymptotic formulas for large $ k
r $ the GF could be used.  Below we find the asymptotic
expressions for GF (\ref{eq26}) at $ r \to \infty $. For real $
\epsilon = E $ the equality
\begin{equation}
{{\epsilon }_{1,2}}\left( k_{\pm }^{\left( 1,2 \right)},f
\right)=E
\label{eq32}
\end{equation}
gives two branches of the electron energy spectrum.  The positive
roots of Eq.(\ref{eq32}) describe the isoenergetic contours $
k=k_{\pm }^{\left( j \right)}\left( E,f \right) $  corresponding
physical electron states in k - space for given energy E. If
$E>{{E}_{0}}$, the roots $ k_{+}^{\left( 1,2 \right)}>0 $ for any
values of f, while roots $ k_{-}^{\left( 1,2 \right)}<0 $. For
$E<{{E}_{0}}$ reals roots of  equation (\ref{eq32}) exist, if
inequality $ \frac{2m({{E}_{0}}-E)}{{{\hbar }^{2}}} \le {{\left(
{{\lambda }^{\left( j \right)}} \right)}^{2}} $ is hold. Both
roots $ k_{\pm }^{\left( j \right)} $ take positive values for the
angles $f$ in which ${{\lambda }_{j}}>0$. Below we will not
consider values of energies $E$  for which the GF exponentially
decreases with coordinate $r$assuming $k_{\pm }^{\left( j
\right)},E\in \mathbb{R}$. At first we substitute asymptotic
expansions for $Si\left( z \right)$ and $Ci\left( z \right)$ at
the large z  (see, for example \cite{bib15}) to the function $
F\left( {{k}_{0}},r \right) $ (\ref{eq23}). At large $r\to \infty
$ and real ${{k}_{0}}$ the main term of expansion reads as
\begin{equation}
F\left( {{k}_{0}}\pm i0,r \right)\approx \frac{i\pi
}{2}{{e}^{i{{k}_{0}}r}}\left[ \left( 1+sign ({{k}_{0}})
\right)\left( sign \left( r \right)\pm 1 \right) \right]+{\mathrm
O}\left( \frac{1}{\left| {{k}_{0}}r \right|} \right);\quad \left|
{{k}_{0}}r \right|\gg 1. \label{eq33}
\end{equation}
As the second step one derive the asymptotic formula for
(\ref{eq23}) by the stationary phase method \cite{bib17}.
Stationary phase points $ f=f_{st}^{\left( j \right)} $ must be
found from equation
\begin{equation}
\begin{aligned}
  & {{\left. \frac{d}{df}\left( k_{\pm }^{\left( j \right)}\cos (f-{{\varphi }_{r}}) \right) \right|}_{f=f_{st}^{\left( j \right)}}}= \\
 & \dot{k}_{\pm }^{\left( j \right)}\cos (f-{{\varphi }_{r}}){{\left. -k_{\pm }^{\left( j \right)}\sin (f-{{\varphi }_{r}}) \right|}_{f=f_{st}^{\left( j \right)}}}=0, \\
\end{aligned}
\label{eq34}
\end{equation}
which leads to the condition $\mathbf{r}\parallel
{{\mathbf{n}}_{v}}$, where ${{\mathbf{n}}_{v}}$ is the unit vector
along the electron velocity ${{\mathbf{v}}^{\left( j
\right)}}={{\nabla }_{\mathbf{k}}}{{\epsilon }_{j}}/\hbar $  (see
also \cite{bib18,bib19})
\begin{equation}
{{\left. \mathbf{r}{{\mathbf{n}}_{v}} \right|}_{f=f_{st}^{\left( j
\right)}}}=r;\quad {{\mathbf{n}}_{v}}\left( f
\right)=\frac{{{\mathbf{v}}^{\left( j \right)}}}{\left|
{{\mathbf{v}}^{\left( j \right)}} \right|}=\mp \left(
-\frac{\dot{k}_{\pm }^{\left( j \right)}\sin f+k_{\pm }^{\left( j
\right)}\cos f}{\sqrt{k_{\pm }^{\left( j \right)2}+\dot{k}_{\pm
}^{\left( j \right)2}}},\frac{\dot{k}_{\pm }^{\left( j
\right)}\cos f-k_{\pm }^{\left( j \right)}\sin f}{\sqrt{k_{\pm
}^{\left( j \right)2}+\dot{k}_{\pm }^{\left( j \right)2}}}
\right).
\label{eq35}
\end{equation}
Here and in all formulas below the point above functions denotes
the derivative on angle f.

As the result of standard calculations we find the asymptotic of
the GF (\ref{eq26})
\begin{equation}
\begin{aligned}
  & G(\epsilon ,\mathbf{r})\simeq -\frac{i}{2\sqrt{2\pi }}\exp \left[ i({{k}_{x0}}\cos {{\varphi }_{r}}+{{k}_{y0}}sin{{\varphi }_{r}})r \right]\times  \\
 & \sum\limits_{j=1,2}{\sum\limits_{s}{\frac{\left( 1+{{\sigma }_{x}}\cos {{\theta }_{j}}+{{\sigma }_{y}}\sin {{\theta }_{j}} \right)}{\hbar {{v}^{\left( j \right)}}\sqrt{\left| {{K}_{j}} \right|r}}}}{{\left. exp\left[ i{{S}_{j}}r\mp \frac{i\pi }{4}sign{{K}_{j}} \right] \right|}_{f=f_{st}^{\left( j \right)}}}+O\left( \frac{1}{r} \right);\quad r\to \infty , \\
\end{aligned}
\label{eq36}
\end{equation}
\begin{equation}
{{S}_{j}}\left( f \right)=k_{\pm }^{\left( j \right)}\left( f
\right)\cos (f-{{\varphi }_{r}}),
\label{eq37}
\end{equation}
\begin{equation}
{{\ddot{S}}_{j}}\left( f_{st}^{\left( j \right)} \right)=\mp
{{K}_{j}}\left( f_{st}^{\left( j \right)} \right)\left(
\dot{k}_{\pm }^{\left( j \right)}{{\left( f_{st}^{\left( j
\right)} \right)}^{2}}+k_{\pm }^{\left( j
\right)}{{(f_{st}^{\left( j \right)})}^{2}} \right).
\label{eq38}
\end{equation}
We assume $ {{S}_{j}}\left( f \right)\in \mathbb{R}$, $ r>0 $, $
{{S}_{j}}\left( f_{st}^{\left( j \right)} \right)\ne 0 $ , $
{{\ddot{S}}_{j}}\left( f_{st}^{\left( j \right)} \right)\ne 0 $.
All functions in Eq. (\ref{eq36}) are calculated in stationary
phase points $ f=f_{st}^{\left( j \right)} $ for which $
\mathbf{rv}>0 $. Summation over s takes into account the existence
of few solutions of Eq. (\ref{eq34}) (few stationary phase points
$ f_{st}^{\left( 2 \right)}\left( s \right) $ for given direction
of vector $\mathbf{r}$(see Ref. \cite{bib12}). It is possible in
the cases when isoenergetic contour $ k=k_{+}^{\left( 2
\right)}\left( E,f \right) $ is nonconvex. In Eq. (\ref{eq36}), $
{{K}_{1,2}}\left( f \right)\ne 0 $ is the curvature of the
isoenergetic curve $ {{\epsilon }_{1,2}}\left( f \right)=E $,
\begin{equation}
{{K}_{j}}\left( f \right)=\frac{k_{+}^{\left( j \right)}{{\left( f
\right)}^{2}}+2\dot{k}_{+}^{\left( j
\right)}{{(f)}^{2}}-k_{+}^{\left( j
\right)}(f)\ddot{k}_{+}^{\left( j \right)}(f)}{{{\left(
\dot{k}_{+}^{\left( j \right)}\left( f \right)+k_{+}^{\left( j
\right)}{{(f)}^{2}} \right)}^{3/2}}};
\label{eq39}
\end{equation}
\begin{equation}
{{v}_{j}}=\frac{1}{\hbar }\sqrt{{{\left( \frac{\partial {{\epsilon
}_{j}}}{\partial k_{\pm }^{\left( j \right)}}
\right)}^{2}}+\frac{1}{k_{\pm }^{\left( j \right)2}}{{\left(
\frac{\partial {{\epsilon }_{j}}}{\partial f} \right)}^{2}}}
\label{eq40}
\end{equation}
is an absolute value of electron velocity. The Eq. (\ref{eq36})
coincides with the results of Ref.\cite{bib12} in the case of $
B=0 $. We do not adduce the GF for single inflection points, which
can exist on the isoenergetic contour  $ {{\epsilon }_{2}}\left( f
\right)=E $ for certain values of SOI constants.  It can be simply
derived in the same way.

\section{Exact results for special case of equal SOI constants}

Let us consider the special case: $ \alpha =\beta $ and the
magnetic field is directed along the $y=-x$ axis, $
\mathbf{B}=\frac{B}{\sqrt{2}}\left( -1,1,0 \right) $. Under these
conditions the Eq. (\ref{eq10}) can be presented in the form
\begin{equation}
\hat{G}\left( \epsilon ,\mathbf{r} \right)=\frac{1}{2{{(2\pi
)}^{2}}}\sum\limits_{\pm }{\left( {{\sigma }_{0}}\pm
\frac{{{\sigma }_{y}}-{{\sigma }_{x}}}{\sqrt{2}}
\right)\int\limits_{-\infty }^{\infty
}{d\mathbf{k}}}\frac{{{e}^{i\mathbf{kr}}}}{\epsilon -{{\epsilon
}_{\pm }}},
\label{eq41}
\end{equation}
where we introduce new functions of energy dimension
\begin{equation}
{{\epsilon }_{\pm }}=\frac{{{\hbar }^{2}}}{2m}\left[ {{\left(
{{k}_{x}}\pm \frac{\sqrt{2}m\alpha }{{{\hbar }^{2}}}
\right)}^{2}}+{{\left( {{k}_{y}}\pm \frac{\sqrt{2}m\alpha
}{{{\hbar }^{2}}} \right)}^{2}} \right]-\frac{2m{{\alpha
}^{2}}}{{{\hbar }^{2}}}\mp h.
\label{eq42}
\end{equation}
The spin angles (\ref{eq6}), that correspond with parts of the sum
with ${{\epsilon }_{\pm }}$, keep the constant directions
${{\theta }_{+}}=\frac{3\pi }{4}$ and ${{\theta }_{-}}=-\frac{\pi
}{4}$. The Eq. (\ref{eq41}) can be rewritten as
\begin{equation}
\hat{G}\left( \epsilon ,\mathbf{r} \right)=\frac{1}{2}{{\sigma
}_{0}}\left( {{G}_{+}}\left( \epsilon ,\mathbf{r}
\right)+{{G}_{-}}\left( \epsilon ,\mathbf{r} \right)
\right)+\frac{{{\sigma }_{y}}-{{\sigma }_{x}}}{2\sqrt{2}}\left(
{{G}_{+}}\left( \epsilon ,\mathbf{r} \right)-{{G}_{-}}\left(
\epsilon ,\mathbf{r} \right) \right),\quad \epsilon \in
\mathbb{C},
\label{eq43}
\end{equation}
where for the function $ G_{\pm} \left( \epsilon  \right) $ one
obtains
\begin{equation}
\begin{aligned}
  & {{G}_{\pm }}(\epsilon )=\int\limits_{-\infty }^{\infty }{\frac{d{{k}_{x}}d{{k}_{y}}}{{{(2\pi )}^{2}}}}\frac{{{e}^{i\mathbf{kr}}}}{\left( \epsilon +\frac{2m{{\alpha }^{2}}}{{{\hbar }^{2}}}\pm h \right)-\frac{{{\hbar }^{2}}}{2m}\left[ {{\left( {{k}_{x}}\pm \frac{\sqrt{2}m\alpha }{{{\hbar }^{2}}} \right)}^{2}}+{{\left( {{k}_{y}}\pm \frac{\sqrt{2}m\alpha }{{{\hbar }^{2}}} \right)}^{2}} \right]}= \\
 & \exp \left( \pm i\frac{\sqrt{2}m\alpha }{{{\hbar }^{2}}}\left( x+y \right) \right){{G}_{2D}}\left( \epsilon +\frac{2m{{\alpha }^{2}}}{{{\hbar }^{2}}}\mp h \right). \\
\end{aligned}
\label{eq44}
\end{equation}
We point out that $G_{2D} \left( \epsilon ,r \right)$  is
well-known GF of free 2D electrons. Particularly the retarded GF
reads as
\begin{equation}
G_{2D}^{R}\left( {{\epsilon }_{j}},r \right)=-\frac{m}{2{{\hbar
}^{2}}}\left\{ \begin{aligned}
  & iH_{0}^{(1)}\left( \sqrt{2m{{\epsilon }_{j}}}|r|/\hbar  \right);\quad {{\epsilon }_{j}}>0 \\
 & \frac{2}{\pi }{{K}_{0}}\left( \sqrt{2m\left| {{\epsilon }_{j}} \right|}|r|/\hbar  \right);\quad {{\epsilon }_{j}}<0 \\
\end{aligned} \right.
\label{eq45}
\end{equation}
where $ H_{0}^{(n)}(x) $ is the Hankel function and $
{{K}_{0}}\left( x \right) $ is the McDonald function.

\section{Densities of electron states and magnetization}

As an example of our results applications we calculate the
electron density of states $\rho \left( E \right)$ and the density
of vector magnetization $\mathbf{m}\left( E \right)$ at $\alpha
\ne \beta $, which are important characteristics of 2D conducting
system (compare with results of Ref. \cite{bib13}).

Density of states can be found from the relation
\begin{equation}
\rho \left( E \right)=-\frac{1}{\pi }\operatorname{Im}\
\operatorname{Tr}{{\left. \left[ {{{\hat{G}}}^{R}}\left(
E,\mathbf{r} \right) \right]\, \right|}_{\mathbf{r}=0}}.
\label{eq46}
\end{equation}
Substituting  the retarded GF (\ref{eq25}) at $ r=0 $from  Eqs.
(\ref{eq22}), (\ref{eq25}) one can derive electron density of
states,
\begin{equation}
\rho \left( E \right)=\frac{m}{\pi {{\hbar }^{2}}};\quad E\ge
{{E}_{0}},
\label{eq47}
\end{equation}
\begin{equation}
\begin{aligned}
  & \rho \left( E \right)=\sum\limits_{j=1,2}{{{\rho }_{j}}\left( E \right)=}\frac{m}{2{{\pi }^{2}}{{\hbar }^{2}}}\sum\limits_{j=1,2}{\oint{d}f}\Theta \left( {{\lambda }^{\left( j \right)}} \right)\Theta \left( {{\xi }^{\left( j \right)}} \right)\sum\limits_{\pm }{\frac{\pm k_{\pm }^{\left( j \right)}}{k_{\pm }^{\left( j \right)}-k_{\mp }^{\left( j \right)}}}\,= \\
 & =\frac{m}{2{{\pi }^{2}}{{\hbar }^{2}}}\sum\limits_{j=1,2}{\oint{d}f}\frac{{{\lambda }^{\left( j \right)}}}{\sqrt{{{\xi }^{\left( j \right)}}}}\Theta \left( {{\lambda }^{\left( j \right)}} \right)\Theta \left( {{\xi }^{\left( j \right)}} \right);\quad E<{{E}_{0}}, \\
\end{aligned}
\label{eq48}
\end{equation}
where ${{\lambda }^{\left( j \right)}}$ and ${{\xi }^{\left( j
\right)}}$ are defined by Eqs. (\ref{eq17}) and (\ref{eq20}). Note
the importance of the relation between electron energy $E$ and the
energy ${{E}_{0}}$ (\ref{eq18}) of branch touch point. The
Eq.(\ref{eq47}) shows that the density of states is the same as
for free 2D electron gas for energies  $ E\ge {{E}_{0}} $. In
opposite case $E<{{E}_{0}}$ the density of states $ \rho \left( E
\right) $ depends on the magnetic field and constants of SOI. The
features of $ \rho \left( E \right) $ related to minima (steps)
and saddle points (peaks) on the energy surfaces (\ref{eq32}).
These points $\left( {{{\tilde{k}}}_{\nu }},{{f}_{\nu }} \right)$
should be found out of the system of equations
\begin{equation}
\frac{\partial {{\epsilon }_{1,2}}}{\partial
\tilde{k}}=\frac{{{\hbar }^{2}}}{m}\left( \tilde{k}-{{\lambda
}^{\left( 1,2 \right)}}\left( f \right) \right)=0;
\label{eq49}
\end{equation}
\begin{equation}
\frac{\partial {{\epsilon }_{1,2}}}{\partial f}=-\frac{{{\hbar
}^{2}}}{m}\tilde{k}{{\dot{\lambda }}^{\left( 1,2 \right)}}\left( f
\right)=0,
\label{eq50}
\end{equation}
from which
\begin{equation}
{{\tilde{k}}_{\nu }}={{\lambda }^{\left( 1,2 \right)}}\left(
{{f}_{\nu }} \right),\quad {{\dot{\lambda }}^{\left( 1,2
\right)}}\left( {{f}_{\nu }} \right)=0.
\label{eq51}
\end{equation}
It is clear that the Eq.(\ref{eq49}) can be satisfied, if $ {{\xi
}^{\left( j \right)}}=0 $ (see Eq.(\ref{eq20})). Therefore
\begin{equation}
{{\epsilon }_{1,2}}\left( {{{\tilde{k}}}_{\nu }},{{f}_{\nu }}
\right)={{E}_{0}}-\frac{{{\hbar }^{2}}}{2m}{{\left( {{\lambda
}^{\left( 1,2 \right)}}\left( {{f}_{\nu }} \right) \right)}^{2}}.
\label{eq52}
\end{equation}
The energy minima $ {{\epsilon }_{1,2}}\left( {{{\tilde{k}}}_{\nu
}},{{f}_{\nu }} \right)=E_{1,2}^{\min } $ correspond to negative
second derivative $ {{\ddot{\lambda }}^{\left( 1,2 \right)}}\left(
{{f}_{\nu }} \right)<0 $ and saddle points $ {{\epsilon
}_{1,2}}\left( {{{\tilde{k}}}_{\nu }},{{f}_{\nu }}
\right)=E_{1,2}^{sad} $ meet the case $ {{\ddot{\lambda }}^{\left(
1,2 \right)}}\left( {{f}_{\nu }} \right)>0 $.

The electron density should be found by integration over all
available energies below Fermi level ${{E}_{F}}$
\begin{equation}
{{n}_{e}}=\sum\limits_{j=1,2}{\int\limits_{E_{j}^{\min
}}^{{{E}_{F}}}{dE}}{{\rho }_{j}}\left( E \right)=\frac{m}{\pi
{{\hbar }^{2}}}\left[ {{E}_{F}}+\frac{m}{2{{\hbar }^{2}}}\left(
{{\alpha }^{2}}+{{\beta }^{2}} \right) \right];\quad {{E}_{F}}\ge
{{E}_{0}}.
\label{eq53}
\end{equation}
We find density of magnetization using its relations with retarded
GF
\begin{equation}
{{m}_{x,y}}\left( E \right)=-\frac{1}{\pi }\operatorname{Im}\
\operatorname{Tr}{{\left. \left[ {{\sigma
}_{x,y}}{{{\hat{G}}}^{R}}\left( E,\mathbf{r} \right) \right]
\right|}_{\mathbf{r}=0}}.
\label{eq54}
\end{equation}
Substituting the GF (\ref{eq22}) into Eq. (\ref{eq54}) after
calculations similar to carried out above one obtain for $ E \ge
{{E}_{0}} $
\begin{equation}
{{m}_{x,y}}\left( E \right)=0,
\label{eq55}
\end{equation}
that follows from symmetry relations
\begin{equation}
{{\lambda }^{\left( 1,2 \right)}}\left( f-\pi  \right)=-{{\lambda
}^{\left( 2,1 \right)}}\left( f \right);\quad k_{\pm }^{\left( 1,2
\right)}\left( f-\pi  \right)=-k_{\mp }^{\left( 2,1 \right)}\left(
f \right).
\label{eq56}
\end{equation}
At $E<{{E}_{0}}$ the density of magnetization becomes
\begin{equation}
\begin{aligned}
  & {{m}_{x,y}}\left( E \right)=\frac{m}{2{{\pi }^{2}}{{\hbar }^{2}}}\sum\limits_{j=1,2}{\oint{d}f}\left\{ \begin{aligned}
  & \cos {{\theta }_{i}} \\
 & \sin {{\theta }_{i}} \\
\end{aligned} \right\}\Theta \left( {{\lambda }^{\left( j \right)}} \right)\Theta \left( {{\xi }^{\left( j \right)}} \right)\sum\limits_{\pm }{\frac{\pm k_{\pm }^{\left( j \right)}}{k_{\pm }^{\left( j \right)}-k_{\mp }^{\left( j \right)}}}= \\
 & =\frac{m}{2{{\pi }^{2}}{{\hbar }^{2}}}\sum\limits_{j=1,2}{\oint{d}f}\left\{
 \begin{aligned}
  & \cos {{\theta }_{i}} \\
 & \sin {{\theta }_{i}} \\
\end{aligned} \right\}\frac{{{\lambda }^{\left( j \right)}}}{\sqrt{{{\xi }^{\left( j \right)}}}}\Theta \left( {{\lambda }^{\left( j \right)}} \right)\Theta \left( {{\xi }^{\left( j \right)}} \right). \\
\end{aligned}
\label{eq57}
\end{equation}
If Fermi energy $ {{E}_{F}}\ge {{E}_{0}} $ the magnetic moment
corresponds to Pauli's paramagnetism of the free electron gas
without SOI
\begin{equation}
{{M}_{x,y}}=\frac{{{g}^{*}}{{\mu
}_{B}}}{2}\sum\limits_{j=1,2}{\int\limits_{E_{j}^{\min
}}^{{{E}_{F}}}{dE}}{{m}_{x,y;j}}\left( E \right)=\frac{m{{\left(
{{g}^{*}}{{\mu }_{B}} \right)}^{2}}}{4\pi {{\hbar
}^{2}}}{{B}_{x,y}},
\label{eq58}
\end{equation}
and it does not depend on SOI constants. Here $
{{m}_{x,y;j}}\left( E \right) $ are the two items in the sum over
j in Eq. (\ref{eq57}).

\section{Summary}

To some up, the exact and asymptotical expressions for the Green's
function (GF) of 2D noninteracting electron gas with combined
Rashba - Dresselhaus spin-orbit interaction in parallel magnetic
field at zero temperature are derived. We split the GF into two
parts either of which depends only on characteristics of the one
branch of spin-orbit split energy spectrum, Eq. (\ref{eq10}). The
GF in the form of double integral is reduced to the single
integral of trigonometric integral functions, Eq. (\ref{eq26}).
This result should be helpful in numerical computations and in
evaluating asymptotic expressions. We present the asymptotic of GF
for large coordinate values which can be used in quantum
mechanical quasiclassical calculations, Eq. (\ref{eq36}). It is
shown that asymptotic formula depends only on two local
characteristics of energy spectrum (\ref{eq32}): the curvature of
isoenergetic curves  and the electron velocity. For the equal SOI
constants and magnetic field direction along one of the symmetry
axis we express the GF by means of Bessel functions, Eq.
(\ref{eq43}). Although this exact result describes the special
case, it may be used for qualitative analysis of different
problems for other (but close) values of parameters. In the
conclusion we demonstrate a usefulness of our results for
calculation of physical quantities. We find the electron density
of states, Eq. (\ref{eq48}), and density of magnetization, Eq.
(\ref{eq57}). These results allow to obtain the clear
interpretation of peculiarities of the dependencies of mentioned
quantities on the energy and of their appearance conditions. We
believe the results for the electron density (\ref{eq53}) and the
magnetization (\ref{eq58}) for the combined Rashba - Dresselhaus
spin -orbit interaction have been obtain for the first time in
this paper.

\end{document}